\documentclass{IEEEtran}
\usepackage{cite}
\usepackage{amsmath,amssymb,amsfonts,amsthm}
\usepackage{algorithm}
\usepackage{algorithmic}
\usepackage{graphicx}
\usepackage{subcaption}
\usepackage{textcomp}
\usepackage[numbers,sort]{natbib}

\usepackage{customsymb}

\newtheorem{theorem}{Theorem}
\newtheorem{proposition}[theorem]{Proposition}

\theoremstyle{definition}
\newtheorem{definition}{Definition}
\newtheorem{example}{Example}
\newtheorem{assumption}[definition]{Assumption}

\def\BibTeX{{\rm B\kern-.05em{\sc i\kern-.025em b}\kern-.08em
    T\kern-.1667em\lower.7ex\hbox{E}\kern-.125emX}}
\begin{document}

\title{Thompson Sampling for Parameterized Markov Decision Processes with Uninformative Actions}

\author{Michael Gimelfarb and Michael Jong Kim


\thanks{Michael Gimelfarb is with the Department of Mechanical and Industrial Engineering, University of Toronto, Toronto, Ontario M5S 3G8, Canada. (e-mail: mike.gimelfarb@mail.utoronto.ca). }

\thanks{Michael Jong Kim was with the Department of Mechanical and Industrial Engineering, University of Toronto, Toronto, Ontario M5S 3G8, Canada. He is now with the Sauder School of Business, University of British Columbia, Vancouver, British Columbia V6T 1Z2, Canada. (e-mail: mike.kim@sauder.ubc.ca).}}

\maketitle

\allowdisplaybreaks
\begin{abstract}
We study parameterized MDPs (PMDPs) in which the key parameters of interest are unknown and must be learned using Bayesian inference. One key defining feature of such models is the presence of ``uninformative" actions that not provide no information about the unknown parameters. We contribute a set of assumptions for PMDPs under which Thompson sampling guarantees an asymptotically optimal expected regret bound of $O(T^{-1})$, which are easily verified for many classes of problems such as queuing, inventory control, and dynamic pricing. 
\end{abstract}


\section{Introduction}

Parameterized MDPs (PMDPs) are dynamic control problems described by parameters of interest whose values are typically unknown, arising in many applications such as queuing and supply chain control, maintenance, and network design. Such problems can be naturally formulated as POMDPs, in which a learnt posterior distribution over the model parameters is incorporated directly into the system state \cite{afeche2013,araman2009,aviv2005}. However, the corresponding dynamic programming equations are typically computationally or analytically intractable due to the so-called ``curse of dimensionality". To remedy this, a variety of approaches have been proposed to do approximately optimal control including myopic and lookahead policies \citep{harrison2012}, least-squares methods \citep{bertsimas2001,keskin2014}, non-parametric approaches \citep{burnetas2000,besbes2012} and reinforcement learning \citep{chowdhury2021}

One particular approach, which we investigate in this work, is Thompson sampling (TS) \cite{thompson1933,russo2018}, which samples a parameter according to the current posterior distribution at each time step, and then solves the underlying MDP assuming that the sampled parameter is correct. The decoupling of the posterior update and sample phase from the optimization phase makes Thompson sampling a computationally and analytically tractable alternative for solving general PMDPs, and it has been applied in a variety of problems settings, notably multi-armed bandit problems \citep{agrawal2012,russo2014}. However, these models can be classified as purely Bayesian optimization problems since, when stripped of all parameter uncertainty, they are fundamentally single-stage optimization problems. On the other hand, much less is known about the theoretical performance of Thompson sampling for stochastic control problems with parameter uncertainty. 

Our work builds upon the stream of literature on Thompson sampling for PMDPs \citep{kim2017,banjevic2019,osband2013,gopalan2015,johnson2015}. Specifically, the closest work to ours is \citep{kim2017}, which showed that Thompson sampling is asymptotically optimal assuming each action taken reveals some additional information about the unknown parameter(s). On the other hand, many important PMDPs contain ``uninformative" actions that do not reveal information about the parameters, e.g. setting lower inventory levels in an inventory control problem may lead to stock-outs \cite{jain2015}, preventative maintenance provides less statistical information about failures \cite{kim2012}, and so-called ``uninformative prices" in dynamic pricing problems \cite{afeche2013}. To this end, \citet{gopalan2015} showed -- under rather general assumptions -- that the number of instants where sub-optimal actions are chosen scales logarithmically with time, with high \emph{probability}. On the other hand, this paper provides a different set of assumptions and analysis, and contributes more precise regret and learning rate bounds for an important class of PMDPs with uninformative actions that hold in \emph{expectation}. Our work contributes more general assumptions and regret analysis than found in \citet{kim2017}, extending the previous optimal regret guarantees for a broader classes of problems with uninformative actions that are of practical interest.

The remainder of the paper is structured as follows. Section \ref{sec:prelim} defines the Thompson sampling algorithm in the context of PMDPs. Section \ref{sec:main} provides a formal statement of the notion of an ``uninformative" action, a set of general assumptions on the problem structure, and corresponding optimal regret and learning rate bounds for Thompson sampling. These results are proved in Section \ref{sec:proofs}, and empirically validated on three important classes of problems, namely admission control, inventory control and dynamic pricing. 

\section{Preliminaries}
\label{sec:prelim}
\subsection{Parameterized Markov Decision Process}

Decision-making in this paper be summarized as a \emph{parameterized Markov decision process} (PMDP), which consists of a finite set of possible states $\statespace$, a finite set of possible actions or controls $\actionspace$, and a finite set of parameters $\parameterspace$ (e.g. hypotheses) that specify both the reward and the state distribution. Specifically, given a control $\smallaction{t} \in \actionspace$ applied at time $t$ in state $\smallstate{t} \in \statespace$, and a parameter $\parameter \in \parameterspace$, the reward and the next state are sampled according to
\begin{equation*}
    \bigstate{t + 1} \sim \statedist_{\parameter}(\cdot \,|\, \smallstate{t}, \smallaction{t}), \quad \bigreward{t} \sim \rewarddist_{\parameter}(\cdot \,|\, \smallstate{t}, \smallaction{t}).
\end{equation*}
We also define the reward function as $\reward{\parameter}{\smallstate{}}{\smallaction{}} \defeq \EL{\parameter}{\bigreward{} | \smallstate{}, \smallaction{}} = \int r \rewarddist_\parameter(r | \smallstate{}, \smallaction{}) \,\mathrm{d}r$, which is assumed to be uniformly bounded on $\mathcal{S} \times \mathcal{A}$.

An \emph{admissible policy} $\mu$ is defined as a sequence of mappings $\largeset{\mu_t}{t \geq 0}$, where $\mu_t(\history{t}) \mapsto \mathscr{P}(\actionspace)$ maps the history of information observed up to time $t$, $\history{t} = \finiteset{\bigstate{0}, \bigaction{0}, \bigreward{0}, \dots \bigstate{t-1}, \bigaction{t-1}, \dots \bigreward{t-1}, \bigstate{t}}$, to a probabilitity distribution over the action space, and $\Pi$ be the set of all admissible policies. On the other hand, a \emph{Markov policy} is a sequence of mappings $\mu_t : \statespace \to \actionspace$; it is further a \emph{stationary policy} if $\mu_0 = \mu_1 = \dots$. Let $\Pi_S$ denote the set of all stationary policies. Given a fixed parameter $\parameter \in \parameterspace$, the goal is to maximize over all $\mu \in \Pi$, the long-run expected average reward starting from an initial state $\bigstate{0} = \smallstate{0}$, given as
\begin{equation}
\label{eqn:optimal_value}
    J_\theta^\mu(\smallstate{0}) \defeq \limsup_{T \to \infty} \frac{1}{T} \EL{\parameter}{\sum_{t=0}^{T-1} \reward{\parameter}{\bigstate{t}}{\bigaction{t}}},
\end{equation}
in which the expectation is computed with respect to $\statedist_\parameter$ and $\bigaction{t}$ are sampled according to $\mu$. We assume that an optimal policy $\mu_\parameter^*$ exists, which must necessarily be a stationary Markov policy. Furthermore, the state process $\mathscr{\bigstate{}}_\parameter = \largeset{\bigstate{t}}{t \geq 0}$ induced by $\mu_\parameter^*$ constitutes a time-homogeneous Markov chain, and we assume that
\begin{assumption}
\label{a1}
    The state process $\mathscr{\bigstate{}}_\parameter = \largeset{\bigstate{t}}{t\geq 0}$ induced by $\mu_\parameter^*$ is ergodic unichain.
\end{assumption}
\noindent Please note that our assumption is weaker than the ergodicity required in \cite{kim2017}, although the main theoretical results therein also follow through under the weaker assumption above \cite{huang1976}.

\subsection{Thompson Sampling}

The decision maker accounts for uncertainty in $\parameter \in \parameterspace$ by modeling it as a random variable $\bigparameter{t}$ in each decision epoch. Starting with \emph{prior distribution} $\prior(\parameter) \defeq \prob{\bigparameter{0} = \parameter \,|\, \history{0}}$, and given history $\history{t}$, the decision maker updates his belief about $\parameter$ by computing the \emph{posterior distribution} using Bayes' theorem:
\begin{equation}
\label{eqn:posterior}
    \posterior{t}(\parameter) \defeq \prob{\bigparameter{t} = \parameter \,|\, \history{t}} = \frac{\mathcal{L}_{\parameter}(\history{t}) \prior(\parameter)}{
		\sum_{\parameteralt \in \parameterspace}{\mathcal{L}_{\parameteralt}(\history{t}) \prior(\parameteralt)}},
\end{equation}
where
\begin{equation*}
    \mathcal{L}_{\parameter}(\history{t}) \defeq \prod_{i=1}^t \rewarddist_\parameter(\bigreward{i-1} \,|\, \bigstate{i-1}, \bigaction{i-1}) \, \statedist_\parameter(\bigstate{i} \,|\, \bigstate{i-1}, \bigaction{i-1})
\end{equation*}
is the \emph{likelihood}.
Note that $\posterior{t}(\parameter)$ is a function of $\history{t}$ and is therefore a random variable. 

The agent selects controls that are consistent with the learned belief $\pi_t$ by following the Thompson sampling algorithm. Formally, this corresponds to an admissible (e.g. history-dependent) policy $\tau$ defined in Algorithm \ref{alg:sampling}.

\begin{algorithm}
	\caption{Thompson Sampling for PMDPs}
	\label{alg:sampling}
	\begin{algorithmic}
		\STATE Initialize $\bigstate{0} \gets \smallstate{0}$ and $\prior$
		\FOR{$t = 0$ to $T$}
		\STATE Sample $\bigparameter{t} \sim \posterior{t}(\cdot)$
		\STATE Solve MDP (\ref{eqn:optimal_value}) with parameter $\bigparameter{t}$ to obtain $\mu_{\bigparameter{t}}^*$
		\STATE Apply control $\bigaction{t} \gets \mu_{\bigparameter{t}}^*(\bigstate{t})$
		\STATE Observe $\bigstate{t+1}$ and $\bigreward{t + 1}$
		\FOR{$\parameter \in \parameterspace$}
		       \STATE $\posterior{t+1}(\parameter) \gets \frac{\rewarddist_{\parameter}(\bigreward{t+1}|\bigstate{t},\bigaction{t}) \statedist_{\parameter}(\bigstate{t+1}|\bigstate{t},\bigaction{t}) \posterior{t}(\parameter)}{\sum_{\parameteralt\in\parameterspace} \rewarddist_{\parameteralt}(\bigreward{t+1}|\bigstate{t},\bigaction{t}) \statedist_{\parameteralt}(\bigstate{t+1}|\bigstate{t},\bigaction{t}) \posterior{t}(\parameteralt)}$ 
		\ENDFOR
		\ENDFOR
		\textbf{end}
	\end{algorithmic}
\end{algorithm}

Intuitively, the posterior update (\ref{eqn:posterior}) should only improve upon the decision maker's belief about $\parameter$ when $\rewarddist_\parameter(\cdot \,|\, \smallstate{}, \smallaction{}) \, \statedist_\parameter(\cdot \,|\, \smallstate{}, \smallaction{})$ is distinguishable for different values of $\parameter$. To see this, we can restate (\ref{eqn:posterior}) as
\begin{equation*}
\posterior{t}(\parameter) 
	= \left[1 + \displaystyle\sum_{\parameteralt\not=\parameter}{
		\frac{\prior(\parameteralt)}{\prior(\parameter)}
		\left(\frac{\mathcal{L}_{\parameteralt}(\history{t})}{\mathcal{L}_{\parameter}(\history{t})} \right)} \right]^{-1},
\end{equation*}
from which we can see that $\posterior{t + 1}(\parameter) = \posterior{t}(\parameter)$ precisely when the ratio of the two likelihoods does not change. Conceptually, the existence of such uninformative state-action pairs $(\smallstate{}, \smallaction{})$ can significantly ``stall'' the posterior updates and prevent the decision maker from learning the correct PMDP parameters. On the other hand, recent analysis of posterior convergence \cite{kim2017} in general PMDPs precludes such scenarios, by imposing rather strong assumptions on the problem structure that do not always hold in practice. In the following section, we will formalize the notion of an uninformative action and provide an alternative and complementary set of assumptions that still provide an asymptotically optimal regret bound.

\section{Thompson Sampling for PMDPs with Uninformative Actions}
\label{sec:main}

\subsection{Uninformative Actions}
\label{subsec:uninformative}

In order to assess the convergence speed of $\posterior{t}$ more precisely, it is necessary to quantify the magnitude of the posterior update. In particular, we define $\nu_{\parameter}(\smallstate{},\smallaction{})$ as the joint probability distribution specified by $\rewarddist_\parameter(\cdot \,|\, \smallstate{}, \smallaction{}) \, \statedist_\parameter(\cdot \,|\, \smallstate{}, \smallaction{})$ and the \emph{KL-divergence}
\begin{equation*}
    \entropy{\nu_\parameter(\smallstate{},\smallaction{})}{\nu_\parameteralt(\smallstate{},\smallaction{})} \defeq \EL{\parameter}{\log\!\left(\frac{\rewarddist_\parameter(\cdot \,|\, \smallstate{}, \smallaction{}) \, \statedist_\parameter(\cdot \,|\, \smallstate{}, \smallaction{})}{\rewarddist_\parameteralt(\cdot \,|\, \smallstate{}, \smallaction{}) \, \statedist_\parameteralt(\cdot \,|\, \smallstate{}, \smallaction{})} \right)}.
\end{equation*}
We require that $\nu_\parameteralt(\smallstate{},\smallaction{})$ is absolutely continuous with respect to $\nu_\parameter(\smallstate{},\smallaction{})$, so that the above quantity is finite.

Formalizing the above intuition, an action $\smallaction{} \in \actionspace$ is called \emph{informative} in state $\smallstate{} \in \statespace$ if, for any two distinct $\parameter, \parameteralt \in \parameterspace$, 
\begin{equation*}
    \entropy{\nu_\parameter(\smallstate{},\smallaction{})}{\nu_\parameteralt(\smallstate{},\smallaction{})} > 0.
\end{equation*}

\begin{assumption}
\label{a3}
    There exists $\specials \in \statespace$ such that $\optimalpolicy{\parameter}{\specials}$ is an informative action in state $\specials$ for all $\parameter \in \parameterspace$.
\end{assumption}

\noindent Note that Assumption \ref{a3} is easy to check in practice since, in the context of Algorithm \ref{alg:sampling}, the policies $\mu_\parameter^*$ are typically computed in advance and cached. Furthermore, it naturally partitions the state-action space into two classes. The first class $\Sigma \subset \statespace \times \actionspace$ consists of state-action pairs $(\smallstate{}, \smallaction{})$ for which $\smallaction{}$ is informative in $\smallstate{}$, and is non-empty since it contains $(\specials, \optimalpolicy{\parameter}{\specials})$ for every $\parameter \in \parameterspace$. The second class is comprised of those state-action pairs for which $\smallaction{}$ are \emph{not} informative with respect to $\smallstate{}$. Thus, Assumption \ref{a3} is much weaker than the one required in \cite{kim2017}, where it is simply assumed that $\Sigma^{\mathsf{C}} = \emptyset$. 

Finally, to ensure that the posterior belief converges at the optimal asymptotic rate, it is necessary that the Markov chain induced by Thompson sampling eventually visits an informative state a positive fraction of the time. One way to guarantee this is to ensure that the induced Markov chain under Thompson sampling mixes with respect to the special state $\specials$, which is now known to be informative. More formally, define $\mathcal{I}(\smallstate{}) \subseteq \actionspace$ as the set of controls that are informative in state $\smallstate{}$, and the following random variables for every $\mu \in \Pi$ and $s \in \statespace$:
\begin{equation*}
    \inform{\mu}(t) \defeq \sum_{i=0}^{t-1} \ind{\bigaction{i} \in \mathcal{I}(\bigstate{i})},\quad \inform{\mu}(t; s) \defeq \sum_{i=0}^{t-1} \ind{\bigstate{i} = s}.
\end{equation*} 
Finally, for each $\smallstate{} \in \statespace$, we define $\actionspace^*(s) \defeq \largeset{\smallaction{} \in \actionspace}{\smallaction{} = \mu_{\parameter}^*(s) \mbox{ for some } \parameter \in \parameterspace}$. We then define $\Pi_C \subset \Pi$ to be the set of policies $\mu$ for which $\mu(\smallstate{}) \in \actionspace^*(s)$ holds for all $\smallstate{} \in \statespace$, e.g. the set of policies whose controls are \emph{consistent} with respect to the policy set $\largeset{\mu_\parameter^*}{\parameter \in \parameterspace}$. 

\begin{assumption}
\label{a4}
    There exists a policy $\bar{\mu} \in \Pi_S$, such that the Markov chain induced by $\bar{\mu}$ is ergodic, and $\inform{\mu}(t; \specials) \geq \inform{\bar{\mu}}(t; \specials)$ holds (with probability 1) for all $t$ and $\mu \in \Pi_C$. 
\end{assumption}

\begin{figure}[t!]
    \centering
    \includegraphics[width=0.8\linewidth]{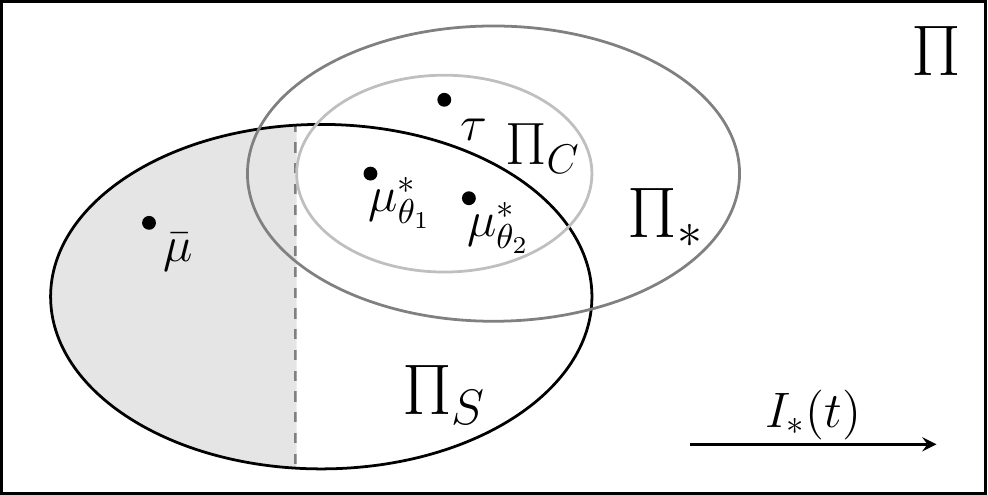}
    \caption{Conceptual illustration of Assumptions \ref{a3} and \ref{a4}.}
    \label{fig:assumptions}
\end{figure}

One way to interpret Assumptions \ref{a3} and \ref{a4} is to define another policy class $\Pi_* \subseteq \Pi$, consisting of all policies $\mu \in \Pi$ for which $\prob{\mu(\specials) \in \mathcal{I}(\specials)} = 1$. Clearly, $\Pi_C \subset \Pi_*$ according to Assumption \ref{a3}, and $\mu_\parameter^* \in \Pi_C$ by construction. Furthermore, by the design of Thompson sampling (Algorithm \ref{alg:sampling}), we can also conclude that $\tau \in \Pi_C$ (although $\tau \not\in \Pi_S$). Finally, Assumption \ref{a4} requires that $\inform{\mu}(t; \specials) \geq \inform{\bar{\mu}}(t; \specials)$ for all ${\mu} \in \Pi_C$, including $\tau$. These facts are summarized conceptually in Fig. \ref{fig:assumptions}. Note that $\bar{\mu}$ can be any policy in the shaded region for which the induced Markov chain is ergodic. As illustrated later in our examples, this policy also often lies in $\Pi_*$, but this is \emph{not} required under our assumptions in general.

In the next section, we will prove that Assumptions \ref{a1}-\ref{a4} are sufficient to achieve asymptotically optimal regret bounds for Thompson sampling with uninformative actions. However, these assumptions are not in general necessary. To see this, suppose Assumption \ref{a4} is relaxed by defining a secondary state $s' \in \statespace$ instead of $\specials$, that need not satisfy Assumption \ref{a3}. Now, if we further assume that there is a non-zero probability of visiting $\specials$ between successive visits to $s'$ under any policy in $\Pi_C$, then states $s'$ and $\specials$ belong in the same communicating class, and the same convergence guarantees also extend to this setting. This generalization would allow us to establish convergence guarantees for learning the service rate parameters in the examples of the following paragraph.

We now provide several important classes of stochastic control problems that satisfy the required assumptions in this paper. Please note, however, that these examples do not satisfy the stronger assumption on the KL-divergence in \cite{kim2017}, and thus the analysis in that paper cannot be applied to them.

\begin{example}[Admission Control]
    At the beginning of each decision epoch, a decision maker decides whether to open or close a server with a fixed capacity $\bar{n} > 0$ (no backlogging of orders is allowed). A single customer arrives in each period with unknown probability $\theta \in (0, 1)$. If the server is open, the customer pays a toll $R \geq 0$ and joins the end of the queue, while if the server is closed, the customer arrival is unobserved. For each customer waiting in the server in each epoch, the decision maker incurs a penalty of $h \geq 0$. Meanwhile at the end of each period, the first customer in queue completes service and leaves with probability $\beta \in (0, 1)$. Opening (closing) the server corresponds to an informative (uninformative) action. It is also easy to check that, if $\beta R \geq h$ (see, e.g., \cite{naor1969}), the optimal policy satisfies $\mu_\parameter^*(0) = ``\mathrm{open}"$ for every $\parameter \in (0, 1)$, e.g. an empty server should be open, and we set $s^* = 0$. Finally, let $\bar{\mu}$ be the policy that always admits a customer unless the server is full. The Markov chain induced by $\bar{\mu}$ is ergodic, and Assumption \ref{a4} holds for any admissible policy $\mu$ that admits when the server is empty. This example can also be generalized to multiple servers, multiple customer types, and general demand or service distributions. \hfill\(\exampleend\)
\end{example}

\begin{example}[Inventory Management]
    A store sells $m$ different types of goods. At the beginning of each decision epoch $t$, the store manager observes the amount of each type of good in stock, $\smallstate{t,i}$, and decides whether or not to fully restock the inventories for each type of good $i$ up to a level $\bar{n}_i > 0$. The delivery time is negligible compared to the length of each decision epoch, and so goods are delivered instantaneously. The wholesale price of a good of type $i$ is $c_i > 0$ and it is sold for $p_i > c_i$. Meanwhile, the cost of holding a good of type $i$ is $h_i > 0$ per item per period. Furthermore, the demand for each good $i$ is modeled as a Poisson random variable with mean $\theta_i > 0$, and stock-outs (e.g. demand exceeding the inventory) are unobserved. Since the manager profits from selling every type of good, it is always optimal to reorder inventory when a particular item is out of stock, and so $\mu_\parameter^*(0, 0, \dots 0)$ requires that all types of goods are restocked. In this case, the inventory level is guaranteed to be positive at the beginning of each epoch and demand for each type of good is observed. Finally, let $\bar{\mu}$ be the policy that chooses to restock every type of good in every decision epoch regardless of the inventory level. Since demand is unbounded, the induced Markov chain is clearly ergodic, and Assumption \ref{a4} holds for all admissible policies $\mu$ that reorder in state $\specials = (0, 0, \dots 0)$. \hfill\(\exampleend\)
\end{example}

\begin{example}[Dynamic Pricing]
    Customer demand in each period follows a Poisson distribution with parameter $\parameter > 0$. When customers arrive, they observe the firm's posted price $p > 0$ and the number of customers already in queue $n \geq 0$, and decide whether to join the queue or leave. Given that the value of the service to the customer is $V > 0$, the waiting cost per epoch is $c > 0$, and the service time per customer is geometric with parameter $\beta \in (0, 1)$, the customer will only join the queue if $V - \frac{c}{\beta}(n + 1) \geq p$ (see, e.g., \cite{borgs2014optimal,chen2001state}). At the beginning of each decision epoch, the firm fixes a price from the set $\finiteset{p_1, p_2, \dots p_m}$, where without loss of generality we may assume $0 < p_1 < p_2 < \dots < p_m < \infty$. We also assume that $p_{m} > V - \frac{c}{\beta}$ so the firm can choose to reject customers, and also assume that $p_1 \leq V - \frac{c}{\beta}$ so the problem is non-trivial. Clearly, the queue has an effective capacity, given by $\bar{n} = \min\largeset{n}{n \geq 0,\, V - \frac{c}{\beta}(n + 1) < p_1}$. Similar to the admission control problem, the cost incurred by the firm is $h > 0$ per customer in queue per period. Finally, let $\specials = 0$ and consider the policy $\bar{\mu}$ that always sets the lowest price $p_1$. The induced Markov chain is ergodic, and Assumption \ref{a4} holds for all admissible policies $\mu$ that fix an attractive price when the queue is empty.
    \hfill\(\exampleend\)
\end{example}

\subsection{Asymptotically Optimal Convergence Rates}

The main theoretical result of this paper establishes that Thompson sampling with uninformative actions achieves asymptotically optimal learning rates (as measured by expected posterior probability of sampling the incorrect parameter) and regret. Here, we sketch out the main reasoning necessary to establish both results, and provide formal proofs in the following sections.

First, it is necessary to understand how fast the time-homogeneous Markov chain induced by the policy $\bar{\mu}$ mixes. In the following section, it will be shown that the Markov chain induced by the policy $\bar{\mu}$ mixes at an exponential (e.g. geometric) rate, and so Assumption \ref{a4} ensures that the special state $\specials$ will be visited at a linear rate asymptotically.

\begin{proposition}
	\label{prop:samplingrateconvergence}
	For $\eta > 0$ sufficiently small, there exists $\delta_{\parameter}, \lambda_{\parameter} > 0$ and $N_{\parameter} \in \naturals$ such that
	\begin{equation*}
	    \probL{\parameter}{\inform{\tau}(t) \geq \eta t} \geq 1 - \delta_{\parameter} \expn{-\lambda_{\parameter} t}, \quad \forall t > N_{\parameter}.
	\end{equation*}
\end{proposition}

By making use of Proposition \ref{prop:samplingrateconvergence}, we then prove that the posterior error converges to zero exponentially fast under our more general assumptions.

\begin{theorem}
	\label{thm:mainA}
	Under Assumptions \ref{a1}-\ref{a4}, and assuming $\pi_0(\theta) > 0$ for all $\theta$, there exist constants $a_{\parameter}$ and $b_{\parameter} > 0$ such that under Thompson sampling,
	\begin{equation*}
	\EL{\parameter}{1 - \posterior{t}(\parameter)} \leq a_{\parameter} \expn{-b_{\parameter} t},\quad \forall t \geq 0.
	\end{equation*}
\end{theorem}

The following regret result follows immediately from Theorem \ref{thm:mainA} proved above, Assumption \ref{a1}, and Theorem 5 in \cite{kim2017}.

\begin{theorem}[\cite{kim2017}]
    \label{thm:mainB}
    Under Assumptions \ref{a1}-\ref{a4}, Thompson sampling achieves a worst-case average regret of $O(T^{-1})$.
\end{theorem}

\section{Proofs of Theoretical Results}
\label{sec:proofs}

\subsection{Proof of Proposition \ref{prop:samplingrateconvergence}}

We first cite the following result.

\begin{proposition}[\cite{glynn2002}]
	\label{prop:Markovconcentration}
	Let $\mathscr{X} = \largeset{X_t}{t\geq 0}$ be a Markov chain taking values in $\mathcal{X}$. Suppose there exists a probability measure $\varphi$ on $\mathcal{X}$, $\lambda > 0$, and integer $m \geq 1$ such that
	\begin{equation}
	\label{eqn:Markovconcentrationcondition}
	\prob{X_m \in \cdot \,|\, X_0 = x} \geq \lambda \varphi(\cdot), \quad \forall x \in \mathcal{X}.
	\end{equation}
	Let $f : \mathcal{X} \to \reals$, and define $Y_i \defeq f(X_i)$ and $C_t \defeq \sum_{i = 0}^{t-1}{Y_i}$, and suppose that $\|f\| = \sup\largeset{\abs{f(x)}}{x \in \mathcal{X}}  < \infty$. Then,
	\begin{equation*}
	\prob{C_t - \E{C_t} \geq t \epsilon} \leq \expn{- \frac{\lambda^2(t \epsilon - 2 \|f\| m / \lambda)^2}{2 t \|f\|^2 m^2}}
	\end{equation*}
	for $t > 2 \|f\| m / (\lambda \epsilon)$.
\end{proposition}

\noindent The condition (\ref{eqn:Markovconcentrationcondition}) is introduced earlier in~\cite{asmussen1992}, in which the authors provide sufficient conditions under which it holds for general Markov chains. In particular, if $\mathcal{X}$ is discrete, then (\ref{eqn:Markovconcentrationcondition}) holds if the $m$-step transition matrix of $\mathscr{X}$ has a column whose elements are uniformly bounded away from zero. Furthermore, if $\mathcal{X}$ is finite, then (\ref{eqn:Markovconcentrationcondition}) holds if $\mathscr{X}$ is aperiodic and irreducible.

Let $\mathscr{\bigstate{}}^{\bar{\mu}} = \largeset{\bigstate{t}^{\bar{\mu}}}{t \geq 0}$ denote the ergodic Markov chain induced by policy $\bar{\mu}$ specified in Assumption \ref{a4}. First, Assumption \ref{a3} states that $\mu_\parameter^*(\specials)$ is an informative action for \emph{every} $\parameter \in \parameterspace$. Since Thompson sampling (randomly) selects \emph{one} of the optimal policies $\mu_{\bigparameter{t}}^*$ in each decision epoch, actions selected by Thompson sampling upon each visit to state $\specials$ must also be informative and thus $\inform{\tau}(t) \geq \inform{\tau}(t; \specials)$. 
By replacing the abstract policy $\mu$ with Thompson sampling in Assumption \ref{a4}, we have $\inform{\tau}(t; \specials) \geq \inform{\bar{\mu}}(t; \specials)$ for all $t \geq 0$, or in other words:
\begin{align*}
\probL{\parameter}{\frac{\inform{\tau}(t)}{t} < \eta} 
&\leq \probL{\parameter}{ \frac{\inform{\tau}(t; \specials)}{t} < \eta } \\
&\leq \probL{\parameter}{ \frac{\inform{\bar{\mu}}(t; \specials)}{t} < \eta },
\end{align*}
which hold for all $\eta > 0$. Therefore, it remains to bound the rightmost quantity.

To this end, let $R(t) = \inform{\bar{\mu}}(t; \specials) / t$. By Assumption \ref{a4}, $\mathscr{\bigstate{}}^{\bar{\mu}}$ is time-homogeneous and ergodic, and hence a limiting distribution exists such that $\EL{\parameter}{R(t)} \rightarrow r$ as $t \to \infty$ for some $r \in (0,1)$. More precisely, for each $\nu > 0$ there exists $N \in \naturals$ (dependent on $\nu$) such that $\abs{ \EL{\parameter}{R(t)} - r} < \nu$ for $t \geq N$, in which case:
\begin{align}
	&\probL{\parameter}{ \abs{R(t) - r } > 2 \nu } \nonumber \\
	&= \probL{\parameter}{ \abs{R(t) - \EL{\parameter}{R(t)} +  \EL{\parameter}{R(t)} - r } > 2 \nu } \nonumber \\
	&\leq \probL{\parameter}{ \abs{R(t) - \EL{\parameter}{R(t)}} 
		+ \abs{ \EL{\parameter}{R(t)} - r } > 2 \nu } \nonumber \\
	&\leq \probL{\parameter}{ \abs{R(t) - \EL{\parameter}{R(t)} } + \nu > 2 \nu } \nonumber \\
	&= \probL{\parameter}{ \abs{R(t) - \EL{\parameter}{R(t)}} > \nu } \nonumber.
\end{align}
Now, choosing any $\eta\in (0,r)$ and substituting $\nu = \frac{r - \eta}{2}$, we obtain:
\begin{align*}
	\probL{\parameter}{R(t) < \eta }
	&= \probL{\parameter}{R(t) - r < - (r - \eta) } \\
	&\leq \probL{\parameter}{ \abs{R(t) - r} > r - \eta} \\
	&\leq \probL{\parameter}{ \abs{R(t) - \EL{\parameter}{R(t)}} > \frac{r - \eta}{2} },
\end{align*}
which is valid for all $t \geq N_1$ where $N_1 \in \naturals$. Finally, we apply Proposition~\ref{prop:Markovconcentration} by setting $Y_i = \ind{\bigstate{i}^{\bar{\mu}} = \specials}$, $C_t = \inform{\bar{\mu}}(t; \specials)$, and $\|f\| = 1$, so that there exists $N_2 \in \naturals$ and $\delta, \lambda > 0$ dependent on $\eta$ and  $\parameter$ such that
\begin{equation*}
	\probL{\parameter}{ \abs{R(t)- \EL{\parameter}{ R(t)}} > \frac{r - \eta}{2} } \leq \delta \expn{- \lambda t}, \quad \forall t \geq N_2.
\end{equation*}
Therefore, by setting $N_\parameter = \max\finiteset{N_1,N_2}$, we have found $\delta_\parameter, \lambda_\parameter > 0$ and $\eta_\parameter \in (0, 1)$ such that
\begin{equation*}
    \probL{\parameter}{\inform{\tau}(t; \specials) \geq \eta_\parameter t} \geq 1 - \delta_\parameter \expn{-\lambda_\parameter t}, \quad \forall t > N_\parameter.
\end{equation*}
This completes the proof. \hfill\(\qedsymbol\)

\subsection{Proof of Theorem \ref{thm:mainA}}

The proof begins in an identical manner to~\cite{kim2017} but diverges after the second paragraph. We begin by writing $\posterior{t}(\parameter)$ as:
\begin{align}
	\label{eqn:posteriorquotient}
	\posterior{t}(\parameter) 
	&= \frac{\mathcal{L}_{\parameter}(\history{t}) \prior(\parameter)}{
		\sum_{\parameteralt \in \parameterspace}{\mathcal{L}_{\parameteralt}(\history{t}) \prior(\parameteralt)}} = \frac{1}{1 + \displaystyle\sum_{\parameteralt\not=\parameter}{c_{\parameteralt} \left(\frac{\mathcal{L}_{\parameteralt}(\history{t})}{\mathcal{L}_{\parameter}(\history{t})} \right)}} \nonumber \\
	&= \frac{1}{1 + \displaystyle\sum_{\parameteralt\not=\parameter}{c_{\parameteralt} \expn{-\log{\left( \frac{\mathcal{L}_{\parameter}(\history{t})}{\mathcal{L}_{\parameteralt}(\history{t})} \right)}} }} \nonumber \\
	&= \frac{1}{1 + \displaystyle\sum_{\parameter\not=\parameteralt}{c_{\parameteralt} \expn{- \sum_{s = 0}^t \log{\Lambda_s^{\parameteralt}}}}},
\end{align}
where $c_{\parameteralt} \defeq \prior(\parameteralt)/ \prior(\parameter)$, and where:
\begin{align*}
	\Lambda_0^{\parameteralt} &\defeq 1, \\
	\Lambda_i^{\parameteralt} &\defeq \frac{\rewarddist_{\parameter}( \bigreward{i} \,|\, \bigstate{i - 1}, \bigaction{i-1})\, \statedist_{\parameter}( \bigstate{i} \,|\, \bigstate{i - 1}, \bigaction{i-1})}{ \rewarddist_{\parameteralt}( \bigreward{i} \,|\, \bigstate{i - 1}, \bigaction{i-1}) \, \statedist_{\parameteralt}( \bigstate{i} \,|\, \bigstate{i - 1}, \bigaction{i-1})}, \quad 0 < i \leq t,
\end{align*}
and where the processes $\largeset{\bigstate{t}}{t \geq 0}$ and $\largeset{\bigaction{t}}{t \geq 0}$ are understood to evolve under Thompson sampling.

We define the processes $\mathscr{Z}^{\parameteralt} = \largeset{Z_t^{\parameteralt}}{t \geq 0}$ as
\begin{equation*}
    Z_t^{\parameteralt} \defeq \sum_{i = 0}^t \log{\Lambda_i^{\parameteralt}}, \quad \parameteralt\not=\parameter \in \parameterspace,
\end{equation*}
and for convenience, the filtration $\mathscr{F} = \largeset{ \mathscr{F}_t}{t \geq 0}$ as the sequence of sigma algebras $\mathscr{F}_t \defeq \sigma(\history{t}, \bigaction{t})$., e.g, by concatenating $\bigaction{t}$ to $\history{t}$. We can now separate $\mathscr{Z}^{\parameteralt}$ into a martingale and a predictable process using the Doob Decomposition theorem \citep{klenke2000}:
\begin{align*}
	Z_t^{\parameteralt} 
	&= M_t^{\parameteralt} + P_t^{\parameteralt} \nonumber \\
	&= \sum_{i = 0}^t \left(\log{\Lambda_i^{\parameteralt}} - \EL{\parameter}{\log{\Lambda_i^{\parameteralt}} | \mathscr{F}_{i - 1}} \right) + \sum_{i = 0}^t \EL{\parameter}{\log{\Lambda_i^{\parameteralt}} | \mathscr{F}_{i - 1}},
\end{align*}
where $\mathscr{M}^{\parameteralt} = \largeset{M_t^{\parameteralt}}{t \geq 0}$ is an $\mathscr{F}_t$-martingale under $\measureP_{\parameter}$ by construction, with $M_0^{\parameteralt} = 0$ since $\Lambda_0^{\parameteralt} = 1$.

Since $\bigaction{i}$ is $\mathscr{F}_{i}$-measurable, for every $i \in \finiteset{1,2\dots t}$, $\parameteralt \in \parameterspace$ with $\parameteralt\not=\parameter$, we have:
\begin{align}
	&\EL{\parameter}{\log{\Lambda_i^{\parameteralt}} \,|\, \mathscr{F}_{i - 1}} \nonumber \\
	&= \EL{\parameter}{ \entropy{\nu_{\parameter}(\bigstate{i-1}, \bigaction{i-1})}{ \nu_{\parameteralt}(\bigstate{i-1}, \bigaction{i-1})} \,|\, \mathscr{F}_{i - 1}} \nonumber\\
	&\geq \EL{\parameter}{ \min_{\parameteralt\not=\parameter \in \parameterspace} \entropy{\nu_{\parameter}(\bigstate{i-1},\bigaction{i-1})}{ \nu_{\parameteralt}(\bigstate{i-1},\bigaction{i-1})} \,\Big|\, \mathscr{F}_{i - 1}} \nonumber\\
	&\geq \EL{\parameter}{ \min_{\parameteralt\not= \parameter} \varepsilon(\parameter,\parameteralt) \ind{\bigaction{i-1} \in \mathcal{I}(\bigstate{i-1})} \,\Big|\, \mathscr{F}_{i - 1}} \nonumber\\
	&\geq \varepsilon \ind{\bigaction{i-1} \in \mathcal{I}(\bigstate{i-1})} \nonumber,
\end{align}
where the constants $\varepsilon(\parameter,\parameteralt) > 0$ follow from the definition of an informative action. As a result, $\largeset{P_t^{\parameteralt}}{t \geq 0}$ is a non-decreasing process with $P_0^{\parameteralt} = 0$ and:
\begin{align*}
    P_t^{\parameteralt} 
    &= \sum_{i = 0}^t \EL{\parameter}{\log{\Lambda_i^{\parameteralt}} | \mathscr{F}_{i - 1}} \geq \sum_{i = 1}^t \varepsilon \ind{\bigaction{i-1} \in \mathcal{I}(\bigstate{i-1})} \nonumber \\
    &= \varepsilon I(t),
\end{align*}
where $I(t) \defeq \inform{\tau}(t)$.

We now take expectation of the posterior in~(\ref{eqn:posteriorquotient}):
\begin{align*}
	\EL{\parameter}{\posterior{t}(\parameter)} 
	&= \EL{\parameter}{ \frac{1}{1 + \sum_{\parameteralt\not= \parameter}{c_{\parameteralt} \expn{- Z_t^{\parameteralt}}}}} \\
	&= \EL{\parameter}{ \frac{1}{1 + \sum_{\parameteralt\not=\parameter}{c_{\parameteralt} \expn{- M_t^{\parameteralt} - P_t^{\parameteralt}}}} } \\
	&\geq \EL{\parameter}{ \frac{1}{1 + \sum_{\parameteralt\not= \parameter}{c_{\parameteralt} \expn{- M_t^{\parameteralt} - \varepsilon I(t)}}}}.
\end{align*}
For $\eta \in (0,1)$, we define the event $\mathcal{G}_t^\eta = \lbrace I(t) \geq \eta t \rbrace$ and condition as follows:
\begin{align*}
	&\EL{\parameter}{\posterior{t}(\parameter)} \\
	&\geq \EL{\parameter}{ \frac{1}{1 + \sum_{\parameteralt\not=\parameter}{c_{\parameteralt} \expn{- M_t^{\parameteralt} - \varepsilon I(t)}}} \,\Big|\, \mathcal{G}_t^\eta} \probL{\parameter}{\mathcal{G}_t^\eta} \\
	&\indent+ \EL{\parameter}{ \frac{1}{1 + \sum_{\parameteralt\not= \parameter}{c_{\parameteralt} \expn{- M_t^{\parameteralt} - \varepsilon I(t)}}} \,\Big|\, (\mathcal{G}_t^\eta)^\mathsf{C}} \probL{\parameter}{(\mathcal{G}_t^\eta)^\mathsf{C}}\\
	&\geq \EL{\parameter}{ \frac{1}{1 + \sum_{\parameteralt\not= \parameter}{c_{\parameteralt} \expn{- M_t^{\parameteralt} - \varepsilon I(t)}}} \,\Big|\, \mathcal{G}_t^\eta} \probL{\parameter}{\mathcal{G}_t^\eta} \\
	&\geq \EL{\parameter}{ \frac{1}{1 + \sum_{\parameteralt\not= \parameter}{c_{\parameteralt} \expn{- M_t^{\parameteralt} - \varepsilon \eta t}}} \,\Big|\, \mathcal{G}_t^\eta} \probL{\parameter}{\mathcal{G}_t^\eta}.
\end{align*}

For $\delta \in (0,\varepsilon \eta)$, we define the event $\mathcal{B}_t(\delta) \defeq \bigcap_{\parameteralt\not=\parameter} \eventset{\abs{M_t^{\parameteralt}} \leq \delta t}$ and the scalar $d_\parameter \defeq \frac{1 - \prior(\parameter)}{\prior(\parameter)}$. Then:
\begin{align*}
	&\EL{\parameter,p}{ \frac{1}{1 + \sum_{\parameteralt\not=\parameter}{c_{\parameteralt} \expn{- M_t^{\parameteralt} - \varepsilon \eta t}}} \,\Big|\, \mathcal{G}_t^\eta } \\
	&= \EL{\parameter}{ \frac{1}{1 + \sum{c_{\parameteralt} \expn{- M_t^{\parameteralt} - \varepsilon \eta t}}} \ind{\mathcal{B}_t(\delta)} \,\Big|\, \mathcal{G}_t^\eta } \\
	&\indent+ \EL{\parameter}{ \frac{1}{1 + \sum{c_{\parameter} \expn{- M_t^{\parameteralt}- \varepsilon \eta t}}} \ind{\mathcal{B}_t(\delta)^\mathsf{C}} \,\Big|\, \mathcal{G}_t^\eta } \\
	&\geq \frac{\probL{\parameter}{\mathcal{B}_t(\delta) \,\big|\, \mathcal{G}_t^\eta }}{1 + d_\parameter \expn{-(\epsilon \eta - \delta) t} },
\end{align*}
which yields the lower bound
\begin{align}
	\label{eqn:firstbound}
	\EL{\parameter}{\posterior{t}(\parameter)} \geq \frac{\probL{\parameter}{\mathcal{B}_t(\delta) \,\big|\, \mathcal{G}_t^\eta}}{1 + d_\parameter \expn{-(\epsilon \eta - \delta) t} } \probL{\parameter}{\mathcal{G}_t^\eta}.
\end{align}

Next, we apply DeMorgan's law and the union bound as follows:
\begin{align}
	\label{eqn:probabilitylowerbound}
	\probL{\parameter}{\mathcal{B}_t(\delta) \,\big|\, \mathcal{G}_t^\eta} 
	&= 1 - \probL{\parameter}{\mathcal{B}_t(\delta)^\comp \,\big|\, \mathcal{G}_t^\eta} \nonumber \\
	&= 1 - \probL{\parameter}{\bigcup_{\parameteralt\not=\parameter} \eventset{\abs{M_t^{\parameteralt}} \leq  \delta t}^\comp \,\Big|\, \mathcal{G}_t^\eta} \nonumber \\
	&\geq 1 - \sum_{\parameteralt\not=\parameter} \probL{\parameter}{\abs{M_t^{\parameteralt}} \geq \delta t \,\big|\, \mathcal{G}_t^\eta}.
\end{align}
Using conditional probability, the summand can be further simplified to
\begin{equation*}
\probL{\parameter}{\abs{M_t^{\parameteralt}} > \delta t \,\big|\, \mathcal{G}_t^\eta} \leq \frac{\probL{\parameter}{\abs{M_t^{\parameter}} > \delta t}}{\probL{\parameter}{\mathcal{G}_t^\eta}}.
\end{equation*}
Next, we apply Proposition~\ref{prop:samplingrateconvergence}, so there exists $N$ (dependent on $\eta$ and $\parameter$) such that for $t > N$, $\probL{\parameter}{\mathcal{G}_t^\eta} \geq \frac{1}{2}$ and thus
\begin{equation*}
	\probL{\parameter}{\abs{M_t^{\parameteralt}} > \delta t \,\big|\, \mathcal{G}_t^\eta} \leq 2 \probL{\parameter}{\abs{M_t^{\parameteralt}} > \delta t}.
\end{equation*}

{Since $\entropy{\nu_{\parameter}(\bigstate{i-1}, \bigaction{i-1})}{ \nu_{\parameteralt}(\bigstate{i-1}, \bigaction{i-1})} < \infty$ for all $\parameteralt \not= \parameter$, $\mathscr{M}^{\parameter}$ is an $\mathscr{F}_t$-martingale with bounded increments, and applying Azuma's inequality \citep{klenke2000} obtains:
	\begin{equation*}
		\probL{\parameter}{\abs{M_t^{\parameteralt}} > \delta t \,\big|\, \mathcal{G}_t^\eta} \leq 2 \probL{\parameter}{\abs{M_t^{\parameteralt}} > \delta t} \leq 4 \expn{-\frac{\delta^2}{2 c^2} t},
	\end{equation*}
which holds for $t$ large, $\eta$ small and $\delta < \varepsilon \eta$.} Plugging this into (\ref{eqn:probabilitylowerbound}):
\begin{align*}
	\probL{\parameter}{\mathcal{B}_t(\delta) \,\Big|\, \mathcal{G}_t^\eta} 
	&\geq 1 - \sum_{\parameteralt\not=\parameter} \probL{\parameter}{\abs{M_t^{\parameteralt}} > \delta t \,\big|\, \mathcal{G}_t^\eta} \\
	&\geq 1 - 4 \left(|\parameterspace| - 1\right) \expn{-\frac{\delta^2}{2 c^2} t},
\end{align*}
and therefore (\ref{eqn:firstbound}) leads to:
\begin{align*}
	\EL{\parameter}{\posterior{t}(\parameter)} 
	&\geq \frac{\probL{\parameter}{\mathcal{B}_t(\delta) \,\big|\, \mathcal{G}_t^\eta}}{1 + d_\parameter \expn{-(\epsilon \eta - \delta) t}} \probL{\parameter}{\mathcal{G}_t^\eta} \\
	&\geq \frac{1 - 4 \left(|\parameterspace|  - 1\right) \expn{-\frac{\delta^2}{2 c^2} t}}{1 + d_\parameter \expn{-(\epsilon \eta - \delta) t} } \left( 1 - \delta \expn{\lambda t}\right) \\
	&\geq \frac{1 - \delta_1 \expn{-\lambda_1 t}}{1 + \delta_2 \expn{-\lambda_2 t}}
\end{align*}
for $t$ sufficiently large and some $\delta_1,\delta_2,\lambda_1,\lambda_2 > 0$ dependent on $\parameter$ and $p$. Finally:
\begin{align*}
	\EL{\parameter}{1 - \posterior{t}(\parameter)} 
	&\leq 1 - \frac{1 - \delta_1 \expn{-\lambda_1 t}}{1 + \delta_2 \expn{-\lambda_2 t}}\\
	&= \frac{\delta_2 \expn{-\lambda_2 t} + \delta_1 \expn{-\lambda_1 t}}{1 + \delta_2 \expn{-\lambda_2 t}} \\
	&\leq \delta_2 \expn{-\lambda_2 t} + \delta_1 \expn{-\lambda_1 t} \\
	&\leq \delta_3 \expn{-\lambda_3 t}
\end{align*}
for $t$ sufficiently large and some $\delta_3, \lambda_3 > 0$. It is easy to find $\delta_4, \lambda_4 > 0$ so that $\EL{\parameter}{1 - \posterior{t}(\parameter)} \leq \delta_4 \expn{-\lambda_4 t}$ holds for all $t \geq 0$. This completes the proof. \hfill\(\qedsymbol\)

\section{Numerical Examples}
\label{sec:numeric}

\begin{figure*}[h!]
    \begin{subfigure}[b]{0.275\linewidth}
        \centering
        \includegraphics[width=\linewidth]{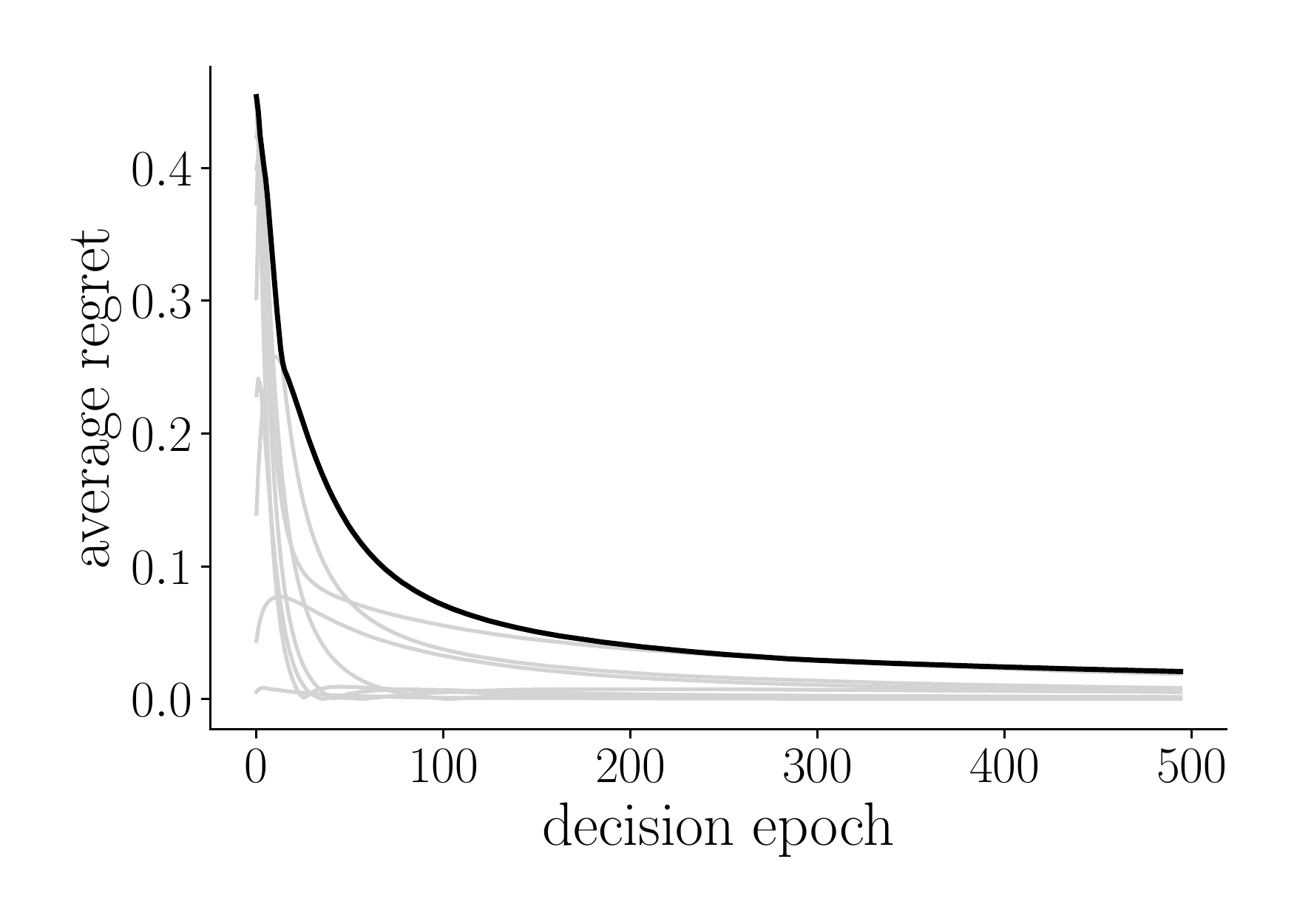}
         \includegraphics[width=\linewidth]{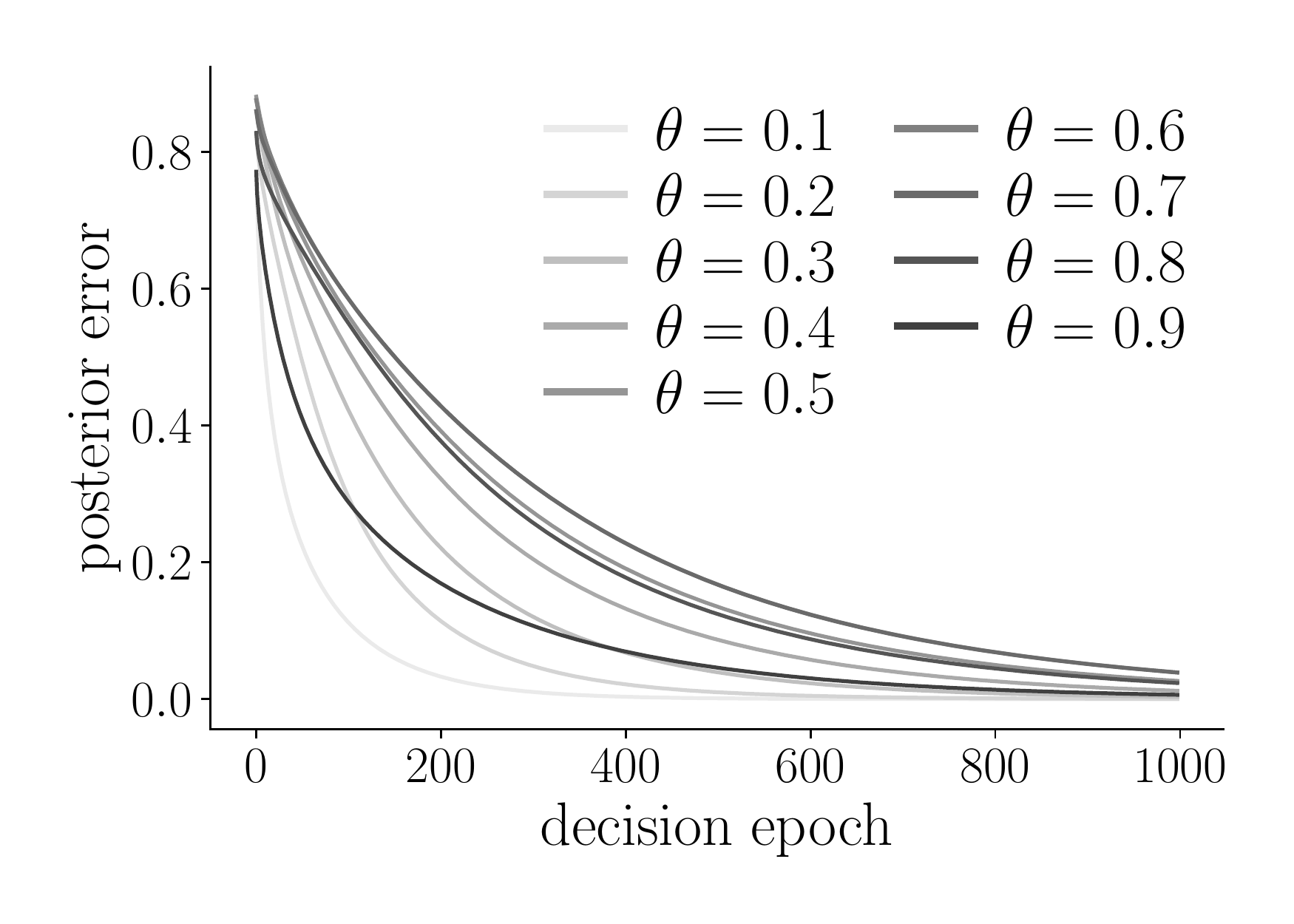}
         \caption{Admission control.}
    \end{subfigure}%
    \begin{subfigure}[b]{0.275\linewidth}
        \centering
        \includegraphics[width=\linewidth]{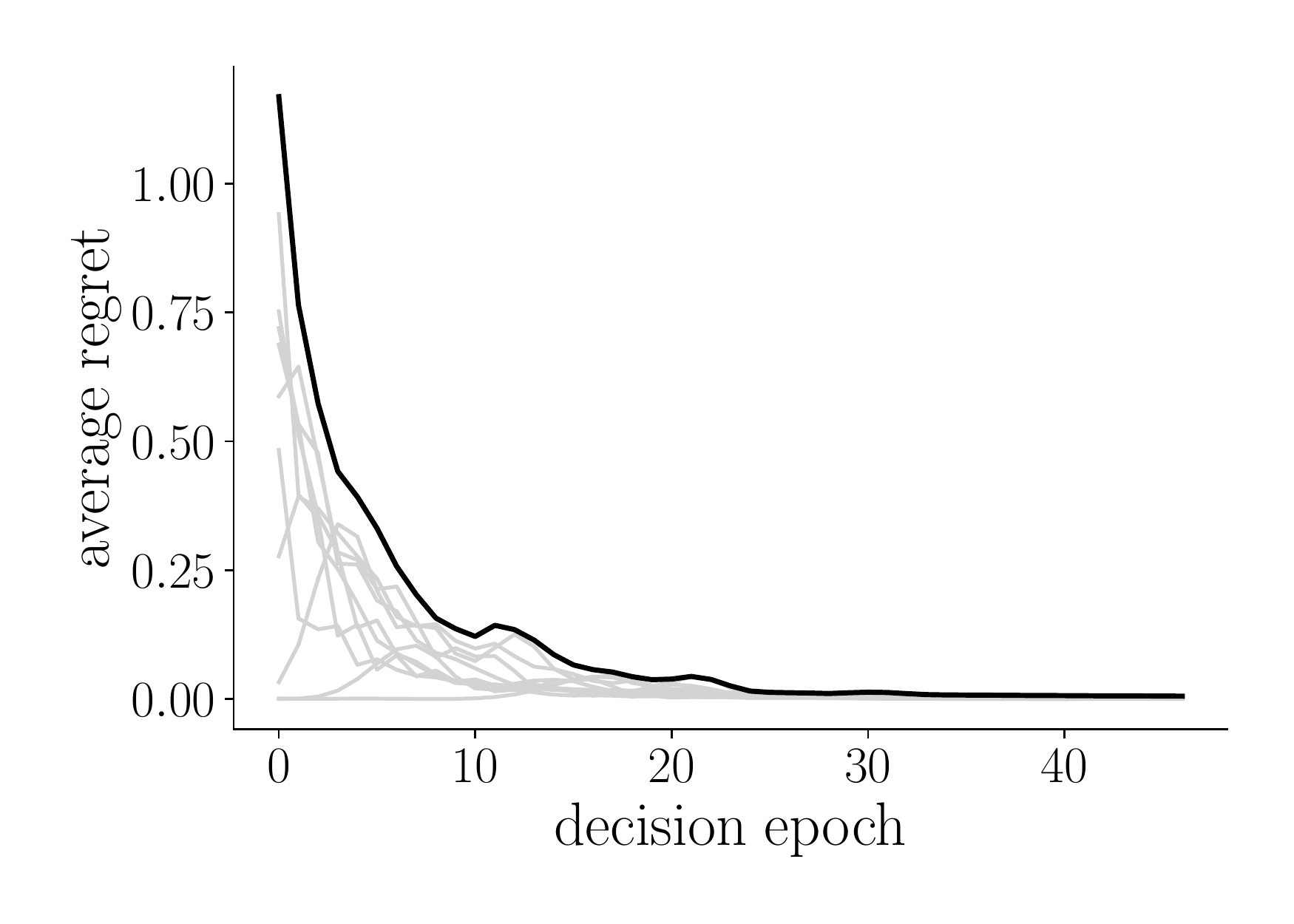}
         \includegraphics[width=\linewidth]{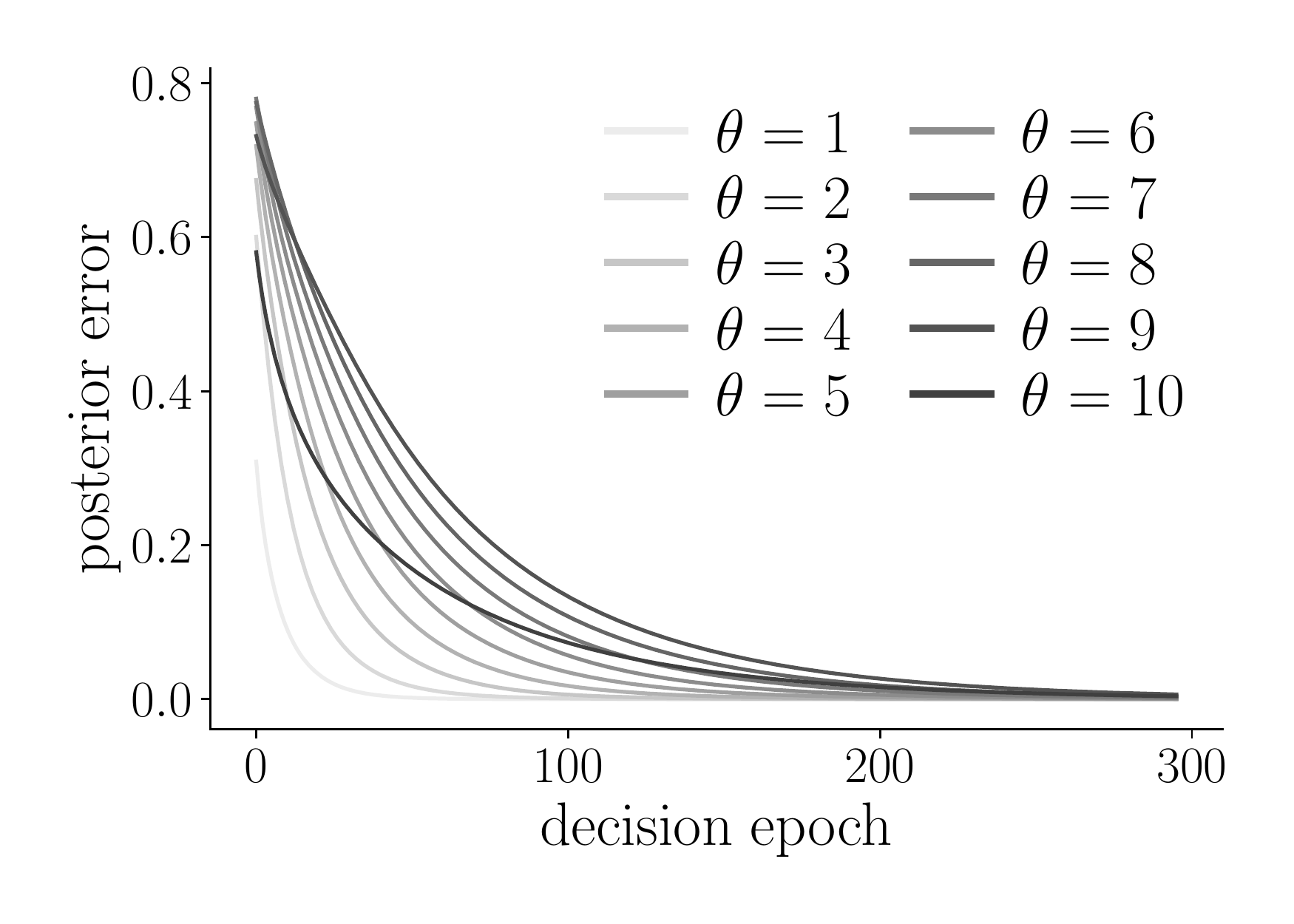}
         \caption{Inventory management.}
    \end{subfigure}%
    \begin{subfigure}[b]{0.275\linewidth}
        \centering
        \includegraphics[width=\linewidth]{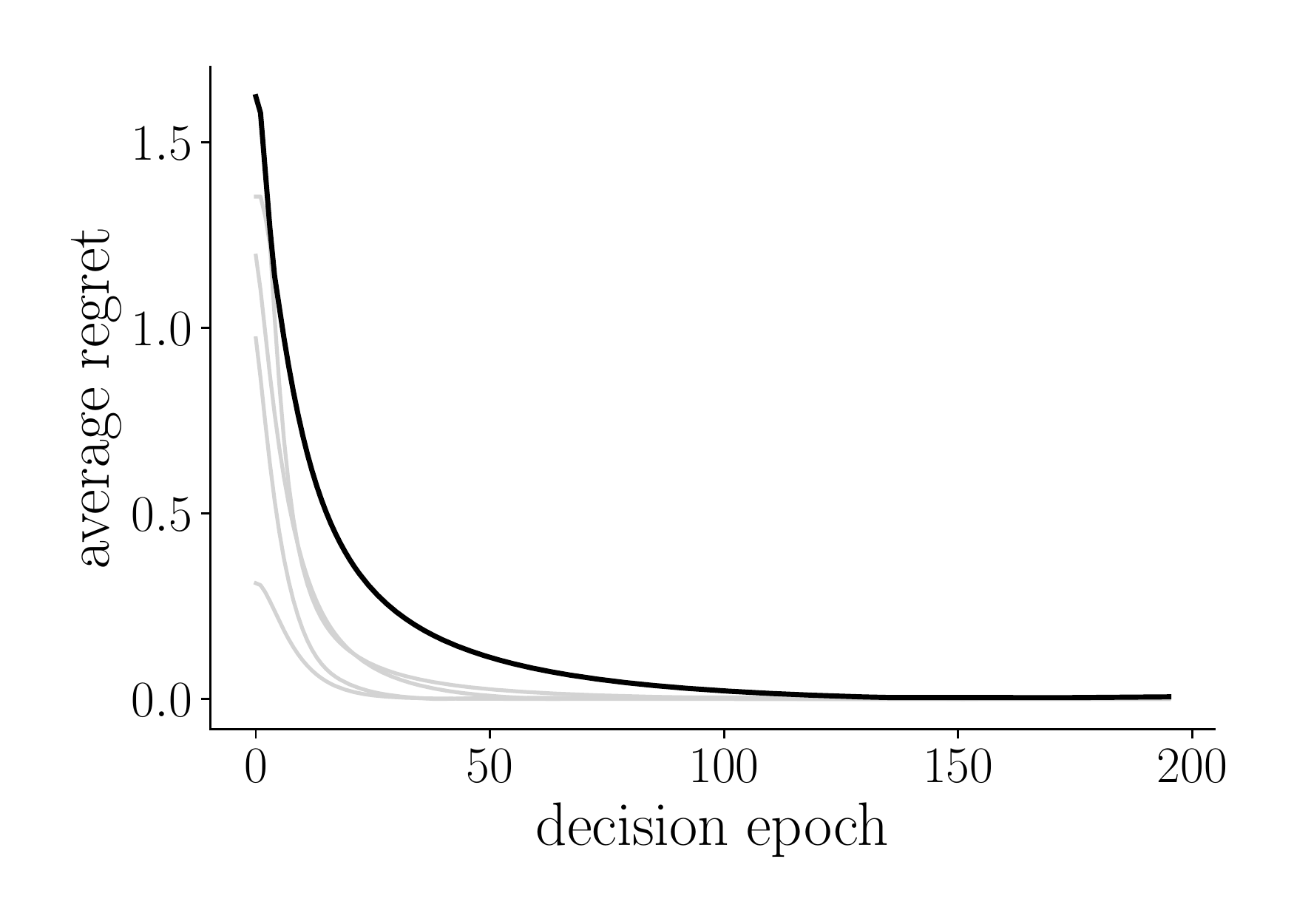}
         \includegraphics[width=\linewidth]{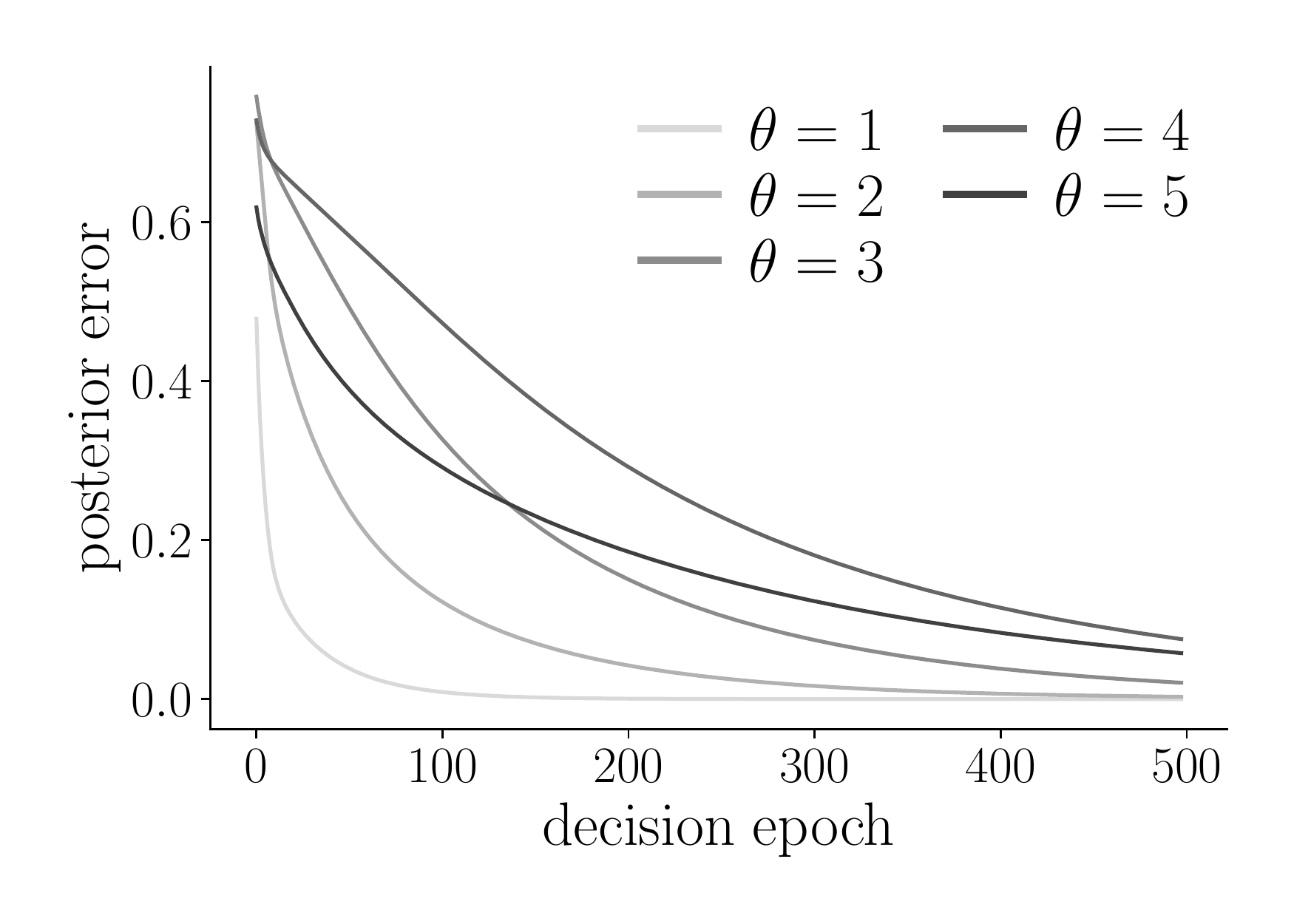}
         \caption{Dynamic pricing.}
    \end{subfigure}%
    \begin{subfigure}[b]{0.175\linewidth}
        \centering
        \includegraphics[width=\linewidth]{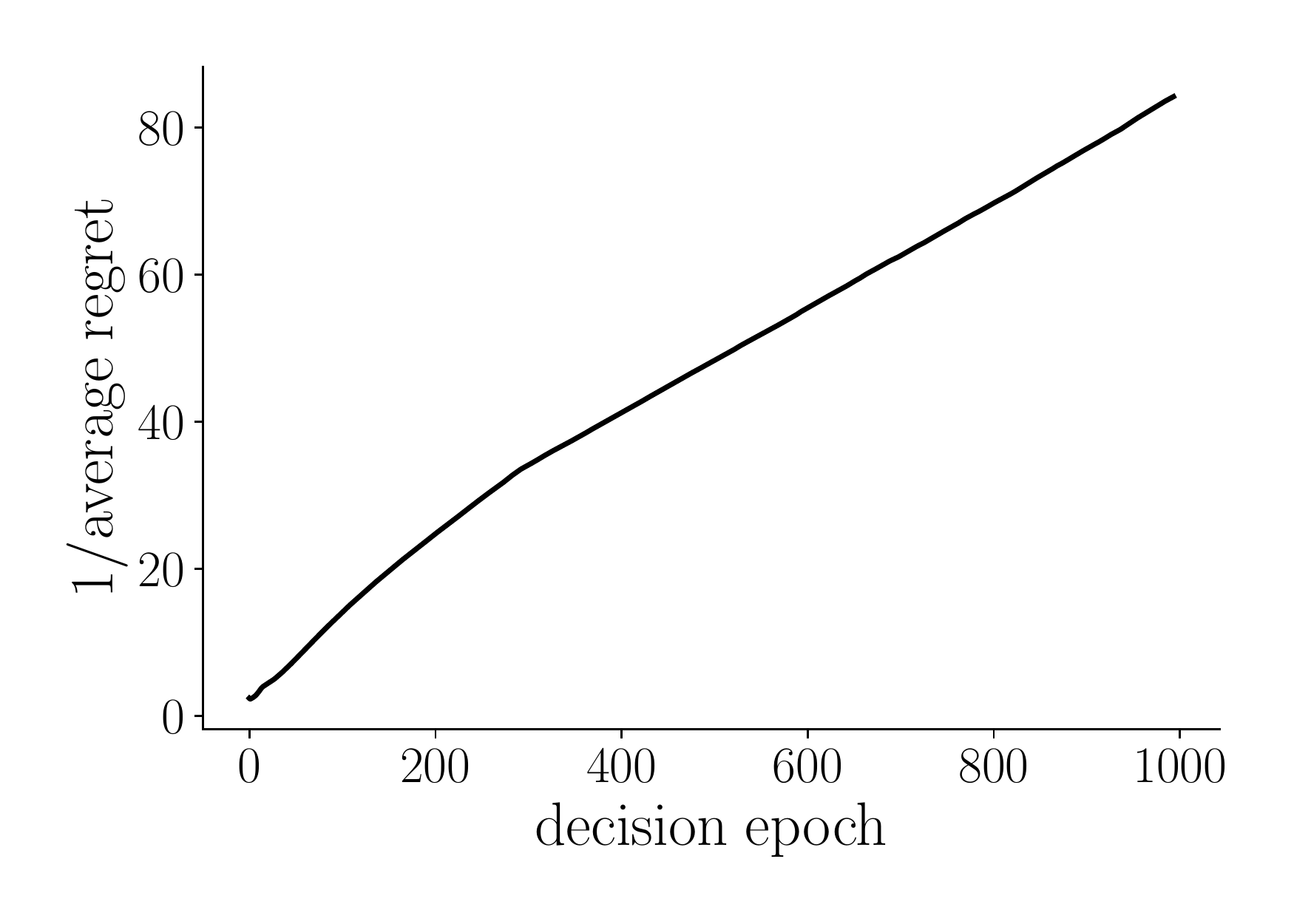}
         \includegraphics[width=\linewidth]{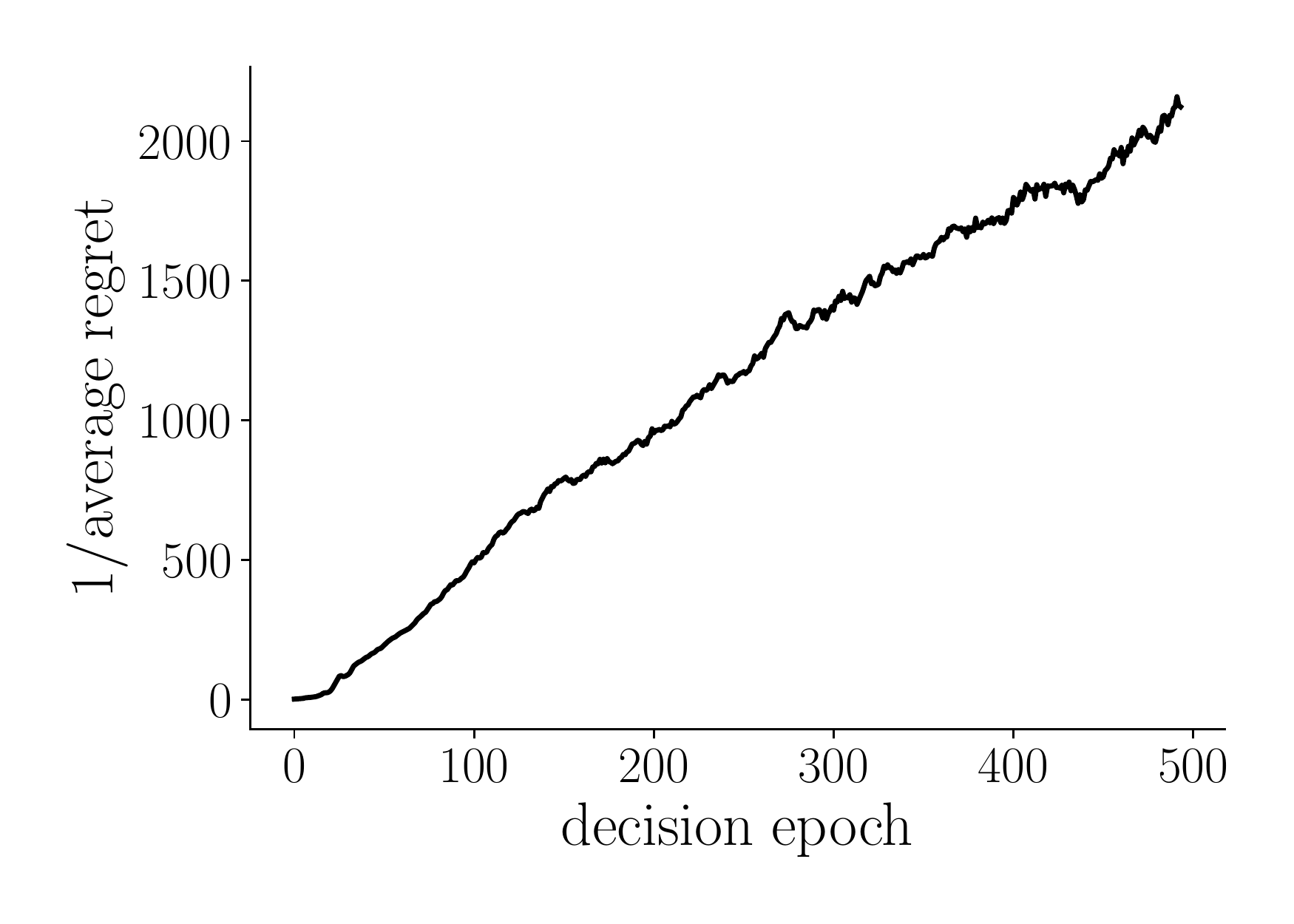}
         \includegraphics[width=\linewidth]{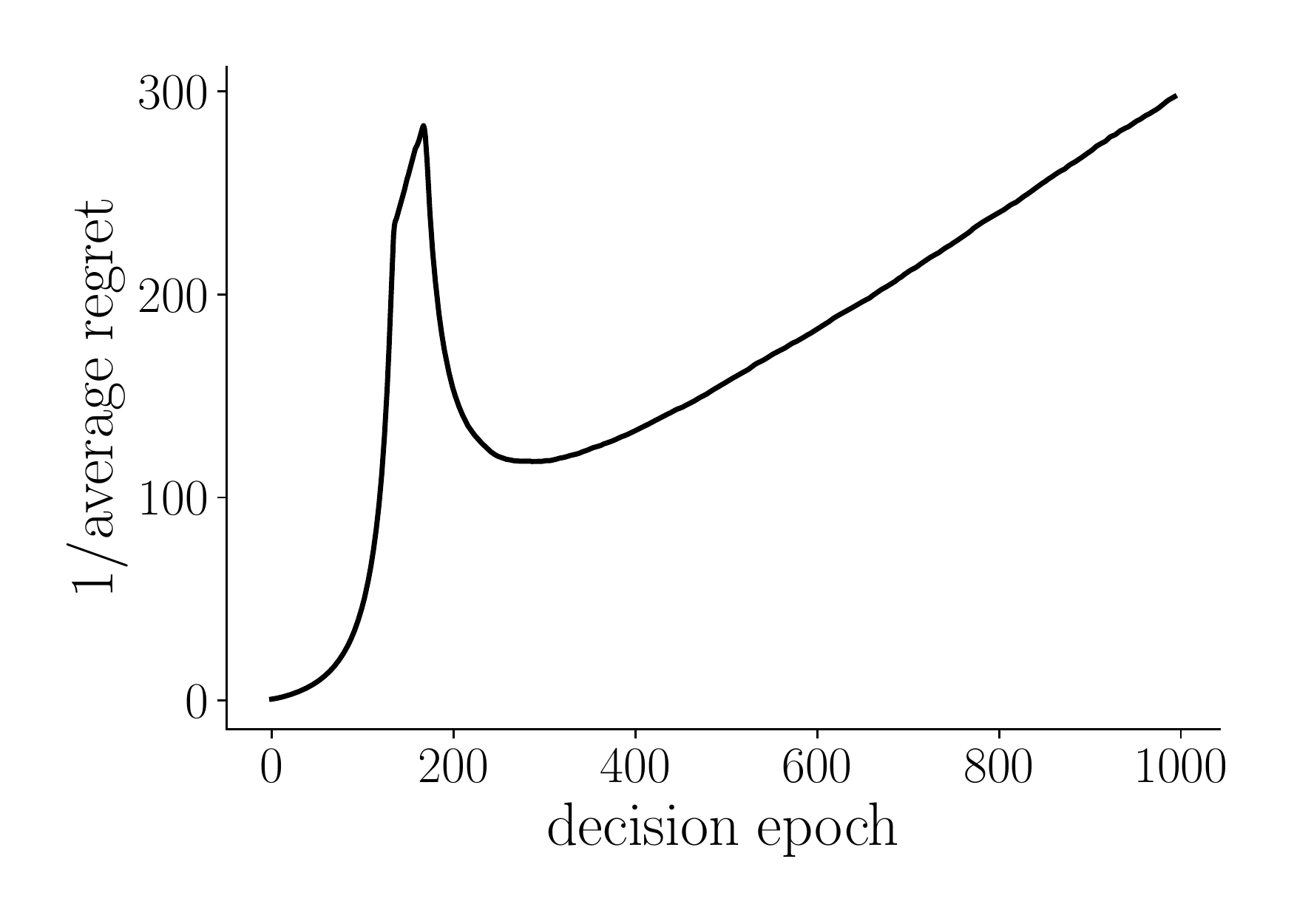}
         \caption{Inverse of the regret.}
    \end{subfigure}
    \caption{Caption} (a)-(c) empirically estimated average regret (top) and posterior error rate for admission control, inventory management and dynamic pricing problems. The dark curve in the topmost figures corresponds to regret of the true unknown parameter value, while the gray curves corresponds to regret of the other parameters. (d) inverse of the empirically estimated regret bounds, showing approximately linear growth for large $T$. 
    \label{fig:main}
\end{figure*}

In this section, we apply Thompson sampling (Algorithm \ref{alg:sampling}) to three instantiations of PMDPs with arrival rate uncertainty and uninformative actions given in Section \ref{subsec:uninformative}. The goal of these experiments is to verify the asymptotically optimal regret bounds hold when Thompson sampling is applied in the online setting with uninformative actions.

The admission control problem assumes $\bar{n} = 40, \,R = 10, \,h = 0.15, \,\beta = 0.3$ and $\parameterspace = \finiteset{0.1, 0.2, 0.3, 0.4, 0.5, 0.6, 0.7, 0.8, 0.9}$. The inventory management problem assumes a single type of good and $\bar{n} = 30, \,c = 2, \,p = 2.8, \,h = 0.01$ and $\parameterspace = \finiteset{1, 2, 3, 4, 5, 6, 7, 8, 9, 10}$. Finally, the dynamic pricing problem assumes $V = 4, \,c = 0.05, \,\beta = 0.5, \,h = 0.01$, prices $p_i = i$ for $i = 0, 1 \dots 5$, and $\parameterspace = \finiteset{1, 2, 3, 4, 5}$. The prior $\prior(\parameter) = 1/|\parameterspace|$ is assigned to a uniform distribution for all problems, and relative value iteration is used to calculate $\mu_\parameter^*$ for each $\parameter$. Finally, to estimate the learning rate and regret for each problem, Thompson sampling is run for $5,\!000,\!000$ sample paths, and the results across paths are averaged for each decision epoch $t$. 

The results are reported in Figure \ref{fig:main}. As illustrated in plots (a)-(c), the empirical expected regret for the true parameter value (shown in black) tends to zero over decision epochs for all problems. To validate that the rate is indeed $O(T^{-1})$, (d) illustrates the corresponding inverse of the empirical regret values, which becomes linear for large $T$ and confirms Theorem \ref{thm:mainB}. The analysis
was conducted on an Intel quad core processor at 2.5
GHz with 8 GB ram, with an average running time of around $10^{-3}$ seconds per sample path and parameter pair. Due to its low time complexity,
Thompson sampling can be easily implemented for larger
parameter spaces and longer planning horizons, which converging at the asymptotically optimal rate.

\section{Conclusion}
We studied parameterized MDPs described by a set of unknown parameters learned using Bayesian inference. A crucial feature of such models was the presence of ``uninformative" actions, which do not provide any information about the unknown parameters and slow down the rate of learning. We contribute a set of assumptions for PMDPs under which Thompson sampling guarantees an asymptotically optimal expected regret bound of $O(T^{-1})$, which are easily verified for many classes of problems such as queuing, inventory control, and dynamic pricing. Numerical experiments validated the theory and showed that, when our assumptions can be verified, provides a computationally efficient algorithm for solving parameterized MDPs.

\appendices

\small
\bibliographystyle{plainnat}
\bibliography{bibl.bib}

\end{document}